\documentclass[%
 reprint,
superscriptaddress,
nofootinbib,
nobibnotes,
 amsmath,amssymb,
]{revtex4-2}

\usepackage{float}
\restylefloat{table}
\usepackage{graphicx}

\usepackage{natbib}

\usepackage{tablefootnote}
\usepackage{xcolor}
\usepackage{ulem}
\usepackage{makecell}
\usepackage{xr-hyper}
\usepackage{hyperref}
\usepackage{cleveref}

\Crefname{equation}{Eq.~}{Eqs.~}
\Crefname{figure}{Fig.~}{Figs.~}
\Crefname{section}{Sec.~}{Secs.~}

\makeatletter
\newcommand*{\addFileDependency}[1]{
  \typeout{(#1)}
  \@addtofilelist{#1}
  \IfFileExists{#1}{}{\typeout{No file #1.}}
}
\makeatother
\newcommand*{\myexternaldocument}[1]{%
    \externaldocument{#1}%
    \addFileDependency{#1.tex}%
    \addFileDependency{#1.aux}%
}

\newcommand{\changes}[1]{#1}

\myexternaldocument{hunkler_peter_SI}

\begin{document}


\title{Fast conformational clustering of extensive molecular dynamics simulation data}
\thanks{Copyright 2023 Hunkler, Peter. This article is distributed under a Creative Commons Attribution (CC BY) License.}

\author{Simon Hunkler}
\affiliation{Department of Chemistry, University of Konstanz}

\author{Kay Diederichs}
\affiliation{Department of Chemistry, University of Konstanz}

\author{Oleksandra Kukharenko}
\email{kukharenko@mpip-mainz.mpg.de}
\affiliation{Theory Department, Max Planck Institute for Polymer Research}

\author{Christine Peter}
\email{christine.peter@uni-konstanz.de}
\affiliation{Department of Chemistry, University of Konstanz}

\date{\today}

\begin{abstract}

We present an unsupervised data processing workflow that is specifically designed to obtain a fast conformational clustering of long molecular dynamics simulation trajectories. In this approach we combine two dimensionality reduction algorithms (cc\_analysis and encodermap) with a density-based spatial clustering algorithm (HDBSCAN).
The proposed scheme benefits from the strengths of the three algorithms while avoiding most of the drawbacks of the individual methods. Here the cc\_analysis algorithm is for the first time applied to molecular simulation data. 
Encodermap complements cc\_analysis by providing an efficient way to process and assign large amounts of data to clusters. The main goal of the procedure is to maximize the number of assigned frames of a given trajectory, while keeping a clear conformational identity of the clusters that are found.
In practice we achieve this by using an iterative clustering approach and a tunable root-mean-square-deviation-based criterion in the final cluster assignment. This allows to find clusters of different densities as well as different degrees of structural identity.
With the help of four test systems we illustrate the capability and performance of this clustering workflow: wild-type and thermostable mutant of the Trp-cage protein (TC5b and TC10b), NTL9 and Protein B. Each of these systems poses individual challenges to the scheme, which in total give a nice overview of the advantages, as well as potential difficulties that can arise when using the proposed method.
\end{abstract}

\maketitle

\section{Introduction \label{sec:intro}}

With the ever-growing power of computers over the last decades, researchers in the field of molecular dynamics (MD) have gotten access to increasingly long trajectories and thereby to increasingly large amounts of data. The introduction of supercomputers which are specifically designed to generate MD trajectories (Anton~\citep{ANTON} and Anton~2~\citep{ANTON2}) is only the latest high point in this development.  
Furthermore, new sampling methods~\cite{Kolinski2016, Gao2019} as well as distributed computing projects, such as Folding@home~\citep{foldingathome}, have contributed to a massive increase in generated simulation trajectories. With this increasing amount of data it is essential to have powerful analysis tools to process and understand underlying systems and processes.

There is a rapid increase in application of unsupervised machine learning methods to analyze molecular simulation data. Two of the most used families of analysis techniques are clustering and dimensionality reduction (DR) algorithms. They help to find low-dimensional subspaces in which important aspects of the original data are preserved and to group the data based on a given similarity measure/metric and thereby gain a better overview and understanding. In practice, most of the times clustering and DR methods are used in combination.
The DR algorithms can be roughly divided into: linear methods (the most known are principal component analysis (PCA)~\citep{PCA_origin_1,PCA_origin_2} and time-lagged independent component analysis (TICA)~\citep{tica_origin_1,tica_origin_2}), nonlinear methods (kernel and nonlinear PCA, multidimensional scaling (MDS)~\citep{MDS_original_1,MDS_original_2} and MDS-based methods like sketch-map~\citep{sketchmap}, isomap~\citep{isomap}, diffusion maps~\citep{Diffusion-Maps,Diffusion-Maps-2} or UMAP~\citep{mcinnes2020umap}, etc.) and autoencoder-based approaches like (encodermap~\cite{autoencoder_original,autoencoder_scheme}, time-autoencoder~\cite{Wehmeyer2018}, variational autoencoders~\cite{variational_autoencoder_scheme} and Gaussian mixture variational autoencoders~\cite{Bozkur2020}).
In terms of clustering algorithms, there are again a wide range of different methods: K-Means~\citep{kmeans,kmeans++}, spectral-clustering~\citep{spectral-clustering}, DBSCAN~\citep{DBSCAN}, density-peak clustering~\citep{Density-Peak-Clustering}, CNN-clustering~\citep{CNN-Clustering}, root-mean-square deviation (RMSD) based clustering~\citep{RMSD-Clustering}, neural-networks-based VAMPnets~\cite{VAMPnets}, etc. For a comprehensive overview of unsupervised ML methods commonly used to analyse MD simulation data we refer to Ref.~\citenum{DR_review}.

\begin{figure*}[!ht]
\begin{centering}
\includegraphics[width=0.8\linewidth]{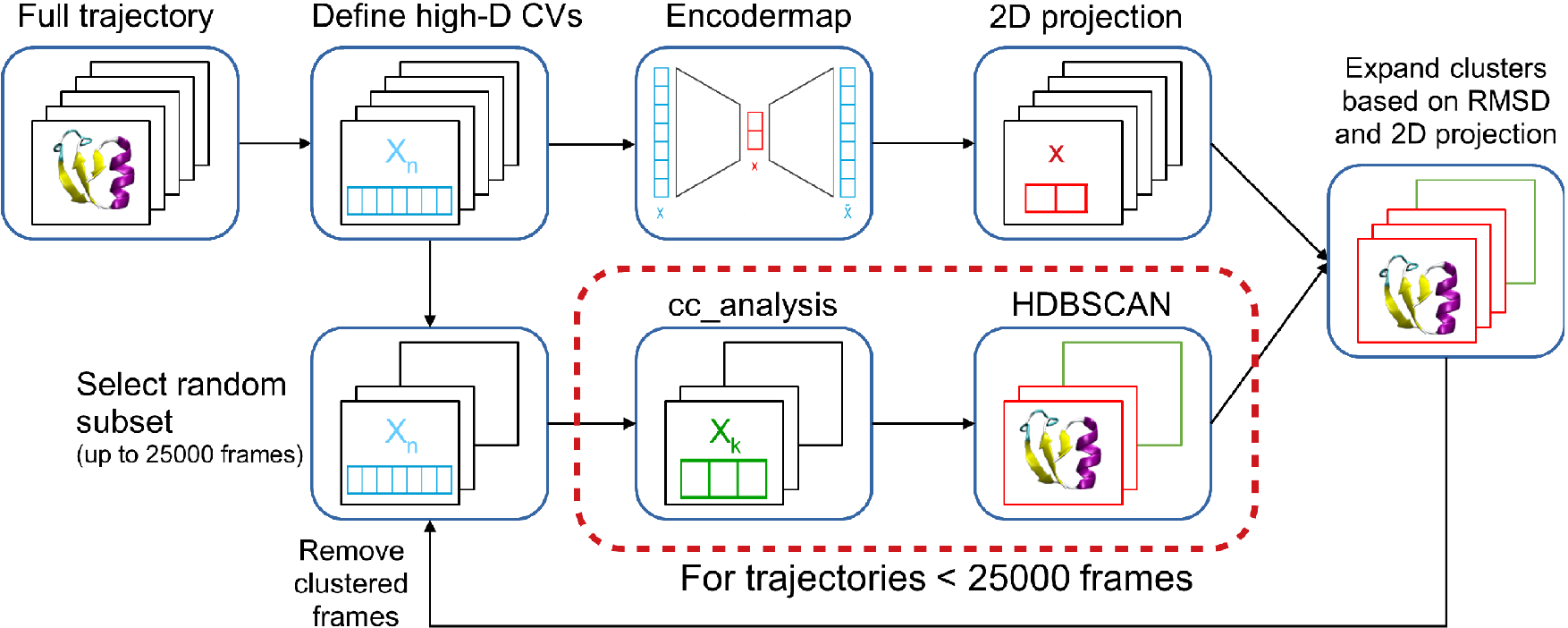}
\caption{\label{fig:Overview-of-the}
Data processing routine presented in this article.}
\end{centering}
\end{figure*}

Even from this incomplete list of available methods it should become obvious that there are a lot of different clustering, as well as DR methods. All these methods have their individual strengths and weaknesses. But there are still open challenges in the successful usage of the listed methods for processing simulation data with high spatial and temporal resolution. This is connected either to the proper choice of hyper-parameters (such as the number of dimensions for DR methods, the number of expected states for some clustering algorithms, neural-networks architectures, different cut-offs, correlation times, etc.), expensive optimisation steps or the amount of data which could be processed simultaneously. In this work we present a data processing scheme which combines three different algorithms in one workflow to create a powerful clustering machinery. It tackles a number of the mentioned challenges as it has a way to define an appropriate lower dimensionality of the data, does not require a priory information about the expected number of states and it is fast in processing extensive MD trajectories with both a very high dimensionality and a large number of observations. It is specifically designed to find conformational clusters in long molecular simulation data (\Cref{fig:Overview-of-the}). 

We use two different DR algorithms, namely an algorithm called ``cc\_analysis'' and the encodermap algorithm. The cc\_analysis method belongs to the family of the MDS-based techniques and was first introduced for the analysis of crystallographic data~\cite{Brehm2014, cc_analysis}. Here it is used for the first time for projecting data of protein conformations. The dimensionality of the cc\_analysis-space which is typically required is more than two (10 to 40 for the systems shown in this work) and the amount of data, which can be efficiently projected simultaneously is limited by the available memory (about 50000 frames for a 72~GB workstation). To process much longer trajectories and to obtain a two-dimensional representation we use the second DR algorithm -- encodermap~\cite{encodermap}. Its loss function however consist of two parts: the autoencoder loss and a MDS-like distance loss, which introduces an interpretability to the resulting 2D representation. Moreover, once the encodermap network is trained, the encoder function can be used to project data to the 2D map in an extremely efficient way. We use encodermap to project data into 2D and for a fast assignment of the additional members to the clusters defined in the cc\_analysis space. Finally we use the HDBSCAN algorithm~\citep{HDBSCAN} to cluster the data in the cc\_analysis space and then visualize the resulting clusters in the 2D encodermap space. 
HDBSCAN is a combination of density and hierarchical clustering, that can work efficiently with clusters of varying density, ignores sparse regions, and requires a minimum number of hyper parameters. We apply it in a non-classical iterative way with varying RMSD-cutoffs to extract the protein conformations of different similarities.

The combination of these three algorithms allows us to leverage their different strengths, while avoiding the drawbacks of the individual methods. Subsequently we will show how the scheme performs on long MD trajectories of wild-type and mutated Trp-cage with native and misfolded meta-stable states (208~$\mu$s and 3.2~$\mu$s long simulations); really extensive simulations of NTL9 (1877~$\mu$s); and Protein B, where only a small percent of the simulation data (5\%) is in the folded state (104~$\mu$s). 

\section{Methods}
\subsection{cc\_analysis}
\label{subsec:DR-Kai}

For dimensionality reduction, we use an cc\_analysis introduced in Ref.~\citenum{Brehm2014, cc_analysis}. 
This algorithm was originally developed to analyse crystallographic data, where presence of noise and missing observations pose a challenge to data processing in certain experimental situations. The method separates the inter-data-set influences of random error from those arising from systematic differences, and reveals the relations between \changes{high-dimensional input features} by representing them as vectors in a low-dimensional space. Due to this \changes{property} we expected it to be highly applicable to protein simulation data, where one seeks to ignore the differences arising from random fluctuations, and to separate the conformations based on systematic differences. In the course of the manuscript we show that this assumption proved to be correct.

The cc\_analysis algorithm belongs to the family of MDS methods~\cite{MDS_original_1}. Its main distinction is that it minimizes the sum of squared differences between Pearson correlation coefficients of pairs of high-dimensional \changes{descriptors} and the scalar product of the  low-dimensional vectors representing them (see \Cref{eq:Kay_MDS}). The procedure places the vectors into a unit sphere within a low-dimensional space. Systematic differences between the high-dimensional \changes{features} lead to differences in the angular directions of the vectors representing them, and purely random differences of data \changes{points} lead to different vector lengths at the same angular direction. The algorithm minimizes, e.g. iteratively using L-BFGS~\citep{L-BFGS}, the expression
\begin{equation}
\Phi(\mathbf{x})=\sum_{i=1}^{N-1}\sum_{j=i+1}^{N}\left(r_{ij}-x_{i}\cdot x_{j}\right)^{2}\label{eq:Kay_MDS}
\end{equation}
as a function of $\bf{x}$, the column vector of the N low-dimensional vectors $\it{\{{x_{k}\}}}$.
Here $r_{ij}$ is the correlation coefficient between \changes{descriptors} $i$ and $j$ in the high-dimensional space and $x_{i}\cdot x_{j}$ denotes
the dot product of the unit vectors $x_{i}$ and $x_{j}$ representing the data in the low-dimensional space; $N$ is the number of \changes{observations, e.g. protein conformations}. The output of cc\_analysis is the N low-dimensional vectors $\it{\{{x_{k}\}}}$, and the eigenvalues of the $\mathbf{xx}^{T}$ matrix.

To understand why this is a sensible approach, one can think
about obtaining the least squares solution of \Cref{eq:Kay_MDS} algebraically by eigenanalysis of the matrix $\bf{r}=\it{\{{r_{ij}\}}}$. 
In that case one would have to solve
\[ \mathbf{xx}^{T}=\mathbf{r} \]
where $\mathbf{r}$ is the matrix of the correlation coefficients
$r_{ij}$. The $n$ strongest eigenvalue/eigenvector pairs (eigenvectors corresponding to the largest eigenvalues) could then be used to reconstruct the $N$ vectors $x_{i}$, which are located in a $n$-dimensional unit sphere. The systematic differences between the input data are thereby shown by the different angular directions in this low-dimensional sphere. This approximation is meaningful because in general the Pearson correlation coefficient can be written as a dot product between two vectors (\changes{after subtraction of the mean and dividing by the standard deviation to scale the vectors to unit length}) and is equal to the cosine of the angle between them. Hence, in an ideal scenario, $\sum_{i,j}^{N} x_{i}\cdot x_{j}$ can exactly reproduce the high-dimensional correlation coefficient matrix and $\Phi(\mathbf{x})$ in \Cref{eq:Kay_MDS} would be equal to zero. 

The length of the vectors is less important than the angle between them. The latter has an inherent meaning: two \changes{high-dimensional feature vectors} with a correlation coefficient of zero between them would \changes{be projected to unit} vectors at $90^{\circ}$ angles with respect to the origin, and two \changes{feature vectors} with a correlation coefficient of one would have a corresponding angle of zero degrees.

Despite the generality of the cc\_analysis approach, by now it was only applied to crystallographic data~\cite{Gildea_2018, Assmann2020}) and protein sequence grouping~\cite{Su2022}. Here we present a first application of cc\_analysis for protein simulation data analysis.

\subsection{Encodermap}
\label{subsec:encodermap}

\begin{figure}[!htbp]
\begin{centering}
\includegraphics[width=1.0\linewidth]{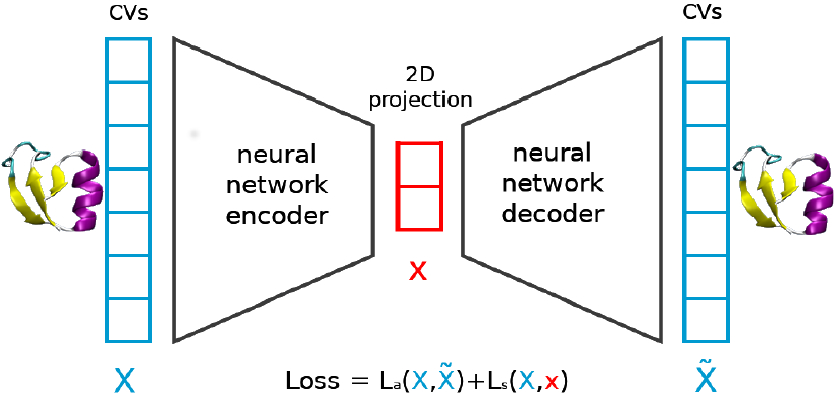}
\caption{\label{fig:encodermap}
Schematic description of encodermap. It has an architecture of the classic autoencoder consisting of two neural networks (encoder and decoder) with the same number of layers and neurons in each layer connected through the bottle-neck layer with two neurons. In addition to autoencoder loss $L_a(X,\tilde{X})$ encodermap loss has a term with the sketch-map loss function $L_s(X,x)$, which improves the quality of two-dimensional projection obtained in the bottle-neck layer (see \Cref{eq:encodermap_loss}).}
\end{centering}
\end{figure}

To accelerate the processing of large datasets, e.g. from extensive simulations, in addition to cc\_analysis, we make use of one more dimensionality reduction technique -- encodermap.  It was developed by~\citet{encodermap} and is used here for fast assignment of data points to clusters as will be explained in details in \Cref{subsec:our_method}.
The method combines the advantages of a neural network autoencoder~\citep{autoencoder_original} with a MDS contribution, here the loss function from 
the sketch-map algorithm~\citep{sketchmap} (\Cref{fig:encodermap}). This combination is exceptionally efficient for projecting large simulation data to the two-dimensional representations: the sketch-map loss function allows to concentrate only on relevant dissimilarities between conformations (ignoring thermal fluctuations and coping with the large dissimilarity values caused by the data's high dimensionality). Furthermore the autoencoder approach allows to avoid complex minimisation steps of the sketch-map projection and to process huge amounts of data in a very short time.

The encodermap loss function $L_{encodermap}$ (\Cref{eq:encodermap_loss}) is a weighted sum of the autoencoder loss $L_{auto}$ (\Cref{eq:auto})  and the sketch-map loss function $L_{sketch}$ (\Cref{eq:sketch}), which emphasizes mid-range distances by transforming all distances \textit{via} a sigmoid function (\Cref{eq:sig}).
\begin{equation}
L_{encodermap}=k_a L_{auto} + k_s L_{sketch} +Reg,
\label{eq:encodermap_loss}
\end{equation} 
\begin{equation}
L_{auto}=\frac{1}{N} \sum_{i=1}^{N}{D(X_i,\tilde{X}_i)},
\label{eq:auto}
\end{equation} 
\begin{equation}
L_{sketch}=\frac{1}{N}\sum_{i\neq j}^N[SIG_{h}(D(X_i,X_j))-SIG_{l}(D(x_i,x_j))]^{2},
\label{eq:sketch}
\end{equation}
where $k_a$, $k_s$ are adjustable weights, $Reg$ is a regularization used to prevent overfitting; $N$ is a number of data points to be projected; $D(\cdot,\cdot)$ is a distance between points, $X$ is a high-dimensional input, $x$ is a low-dimensional projection (the bottleneck layer); $SIG_{h}$ and
$SIG_{l}$ are sigmoid functions of the form shown in \Cref{eq:sig}. 
\begin{equation}
\textrm{SIG}_{\sigma,a,b}(D)=1-(1+(2^{\frac{a}{b}}-1)(\frac{D}{\sigma})^{a})^{-\frac{b}{a}},
\label{eq:sig}
\end{equation}
where $a$, $b$ and $\sigma$ are parameters defining which distances to preserve.

\subsection{Hierarchical Density-Based Spatial Clustering of Applications with Noise (HDBSCAN)} \label{subsec:HDBSCAN-Clustering}

The HDBSCAN~\cite{HDBSCAN, McInnes2017} can be approached from two different sides:
it can be described as a hierarchical implementation of a new formulation
of the original DBSCAN~\citep{DBSCAN} algorithm called DBSCAN{*} by
\citet{HDBSCAN} or it can be formulated as a robust version
of single-linkage clustering with a sophisticated method to obtain
a flat clustering result, as done by \citet{McInnes2017}. 
Here we describe it through the second approach.

\begin{figure}
\begin{centering}
\includegraphics[width=1.0\linewidth]{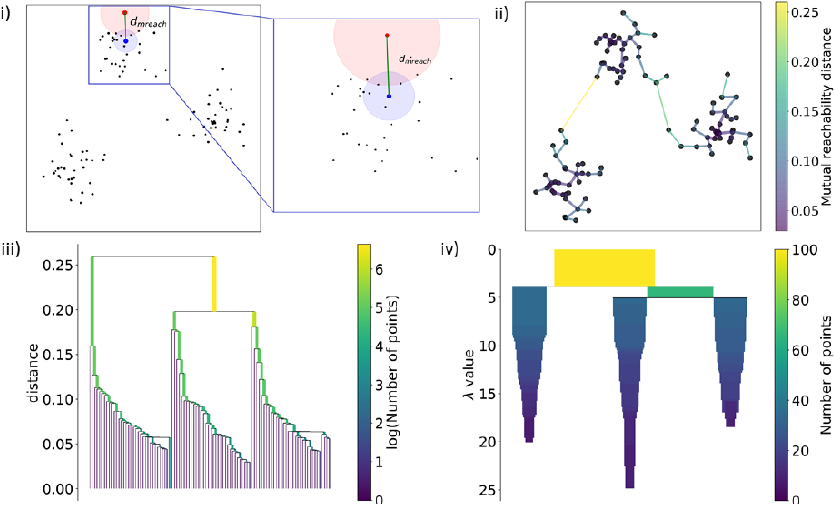}
\caption{\label{fig:HDBSCAN-fig}
Application of HDBSCAN on a toy data set with three clusters. i) Example for the computation of the MRD for two points (red and blue). The red and blue circles indicate the farthest distance to the 5 nearest neighbours for both points. One can see that the distance between the red and blue points (green line) is larger than both the radii of the blue and the red circle. Therefore in this case the green line  distance is chosen as MRD.
ii) The minimum spanning tree based on the MRDs.
iii) The cluster hierarchy. 
iv) The condensed clustering.}
\end{centering}
\end{figure}

In the first step the algorithm introduces the so-called mutual reachability
distance (MRD) (\Cref{eq:mutual}), which transforms the space to make
sparse points even sparser but does not significantly change the distance
between already dense points. 
\begin{equation}\label{eq:mutual}
\begin{split}
D_{mreach-k}(x_i,x_j)&= \\
&max\{core_{k}(x_i),core_{k}(x_j),D(x_i,x_j)\},
\end{split}
\end{equation}
where $x$ are points being clustered, $k$ is a constant which determines a number of nearest neighbouring points, $core_k(x)$ is a function, which finds the maximum distance between a point $x$ and its $k$ nearest neighbours; $D(\cdot, \cdot)$ is a distance between two points. The maximum of three distances is selected as the MRD (\Cref{fig:HDBSCAN-fig} i)).

In the next step the minimum spanning tree based on the MRDs is build via Prim's algorithm~\cite{Prims_alg} (see \Cref{fig:HDBSCAN-fig} ii)). This is done by starting with the
lowest MRD in the data set and connecting the two points by a straight
line. In the following steps always the next nearest data point to
the existing tree, which is not yet connected, is added to the tree.

Once the minimum spanning tree is generated the cluster hierarchy
can be built. This is done by first, sorting the edges of the tree
by distance. Then the algorithm iterates over the edges, always merging
the clusters with the smallest MRD. The result of this procedure can
be seen in \Cref{fig:HDBSCAN-fig} iii).

In order to extract a flat clustering form this hierarchy, a final
step is needed. In this step the cluster hierarchy is condensed down,
by defining a minimum cluster size and checking at each splitting
point if the new forming cluster has at least the same amount of members
as the minimum cluster size. If that is the case, then a new cluster
is accepted, if not then the data points splitting off are considered
noise. The condensed tree of a toy system can be seen in \Cref{fig:HDBSCAN-fig} iv).

\subsection{Introduction of a new clustering workflow}
\label{subsec:our_method}

In this article we present a data processing routine which we found to be extremely efficient for large molecular dynamics simulation trajectories. It relies on the three algorithms introduced above. A schematic description is given in \Cref{fig:Overview-of-the}. In this workflow a given data set is clustered iteratively until either a specified amount of data points are assigned to clusters or a maximum number of iterations have been reached.

\Cref{fig:Overview-of-the} illustrates the sequence of data processing steps along the clustering workflow. 
In the first step a high-dimensional collective variable (CV) is chosen. 
For all systems that are shown in this article all pairwise distances between the $\textrm{C}_{\alpha}$ atoms were selected.
After a CV has been chosen, for trajectories containing more than 25,000 frames, encodermap is trained on all data. Thereby we obtain a function which can be used to project data very efficiently to the newly generated 2D space.
In parallel, a random subset from the entire data set is generated. The reason to use such a subset is a limitation that comes with the cc\_analysis dimensionality reduction. As mentioned in \Cref{subsec:DR-Kai} the cc\_analysis algorithm works with the correlation matrix. This means that the Pearson correlation coefficients of the selected CV (here the pairwise c-alpha distances) are calculated for all unique pairs of frames, and used as input to cc\_analysis.  However the larger a data set is, the larger the correlation coefficient matrix will be, until it is no longer efficient to work with that matrix due to very long computation times as well as memory issues. Therefore a subset is created, by randomly selecting up to 25,000 data points from the entire data set. This subset is then used in the cc\_analysis dimensionality reduction to project the high dimensional CVs (between 190 and 1081 dimensions for the systems in this article) to a lower dimensional subspace (20 to 30 dimensions for the systems in this article). The choice of the appropriate amount of reduced dimensions is done by searching for a spectral gap among the cc\_analysis eigenvalues. Once the cc\_analysis space has been identified, a clustering is generated by applying the HDBSCAN algorithm to that lower dimensional data. A detailed description on how to choose the dimensionality for cc\_analysis and the parameters for HDBSCAN is given in the supporting information (SI), Sec.~S-I.
 
We use two different DR algorithms in the workflow due to the following reasons. For once, the cc\_analysis algorithm is used to project the smaller subsets to a still comparably high-dimensional subspace, which holds more information compared to the 2D projection of encodermap. This higher dimensional subspace is therefore very well suited to be the clustering space. Once the data subset is clustered in the cc\_analysis space, the 2D encodermap space is used to assign the points that were not a part of the subset to the found clusters. The 2D projection is very well suited to do a fast assignment of additional points from the data set, as well as to serve for visualization purposes. Additionally, encodermap is able to project huge data sets very time-efficiently. Hence, the generated 2D projection of a given data set can be used to avoid the main disadvantage of the cc\_analysis algorithm in the way we use the algorithm here, which is having to use subsets of the data due to memory issues.
In order to circumvent this disadvantage, we build a convex hull in the 2D space for each cluster that was found in the cc\_analysis space. If an unassigned point lies inside a convex hull, the RMSD to the central conformation of that cluster is computed. In case the RMSD is inside a given cutoff, the data point is considered to be part of that cluster, else it is not assigned to the cluster. This RMSD cutoff is chosen by taking the weighted mean of all average internal cluster RMSDs~\footnote{By the average internal cluster RMSD we mean the average RMSD of all conformations to the cluster centroid.} of the first clustering iteration. We found that this procedure generates structurally quite well defined clusters with a low internal cluster RMSD since the RMSD criterion is based on well defined conformational states that emerged from cc\_analysis combined with HDBSCAN. However there is also the possibility to identify more fuzzy clusters that only share a general structural motif by using a larger RMSD cutoff for the assignment. An example of the identification of such fuzzy clusters is described in \Cref{subsec:trp-cage}.

By introducing a RMSD criterion in the last step, we force the clustering to only include structurally very similar conformations in the respective clusters. There are of course various other clustering algorithms, which group MD data sets based on their RMSD values, e.g. an implementation~\citep{RMSD-Clustering} in the GROMACS software package~\citep{GROMACS}. Such RMSD-based clustering algorithms have been used in the MD community for at least 20 years now and they are a very obvious choice for conformational clusterings of MD trajectories. They directly compare the positions of specified atoms in various conformations of a molecule and then group the individual conformations along the trajectory using a cutoff value.
However these methods generally rely on the full RMSD matrix of a given data set. For larger trajectories it becomes almost infeasible to compute these matrices due to extremely long computation times as well as memory issues that arise when working with such large matrices. 
In our workflow we can circumvent these issues by only having to compute the RMSD between the coordinates of C$_\alpha$ atoms of the central conformations of each cluster and the data points that lie inside the convex hull of the respective clusters in the 2D space.

In case a given system has less then about 50,000 frames, it is in principle also possible to omit the encodermap part, since the cc\_analysis algorithm is able to handle the entire data set. If this approach is chosen, the user can either entirely skip the RMSD criterion, or the members of clusters that are found in the cc\_analysis space can still be accepted/rejected based on a RMSD cutoff. An advantage of using both the cc\_analysis algorithm and the encodermap algorithm together is the possibility to check the dimensionality reduction steps on the fly. Since the clustering is done in one DR space, but visualized in the other, narrow and well defined clusters in the 2D space indicate that the 2D map separates the different conformational clusters nicely and that therefore the chosen encodermap parameters were well selected.

Our clustering scheme is not very dependent on the quality of encodermap projection, as it is used only to assign additional structures to already identified clusters. Since the clustering itself is done in the higher dimensional cc\_analysis space and the final cluster assignment uses a RMSD cutoff. The only requirement that the scheme poses towards the 2D map is that similar conformations are located close to each other in the map. This is achieved by the MDS-like distance loss part of the overall loss function of encodermap. 

Remaining points which were not assigned to any cluster after one clustering iteration are then used as a new pool of data, from which the new random subset is build. This whole cycle is repeated until a certain amount of data points are assigned to clusters or until a certain number of clustering iterations are performed.
To decide on a stopping point for the iterative procedure we rely on two possible convergence criteria: either a percentage of assigned conformations or average cluster sizes found at an iteration. If we observe a plateau in the percentage of unassigned data points during several successive iterations the clustering procedure is stopped. 
Due to the design of our workflow, the average cluster size of newly added clusters will decrease with each iteration.
Therefore, the average size of newly added clusters or the convergence of this property during successive iterations can also be used as a stopping criterion.
Examples are shown in SI, Sec.~S-II, Fig.~S2.

\begin{table*}[ht]
\begin{tabular}{|c|c|c|c|c|}
\hline 
 & \thead{Trp-cage RE (TC5b) \\ 
 \includegraphics[scale=0.2]{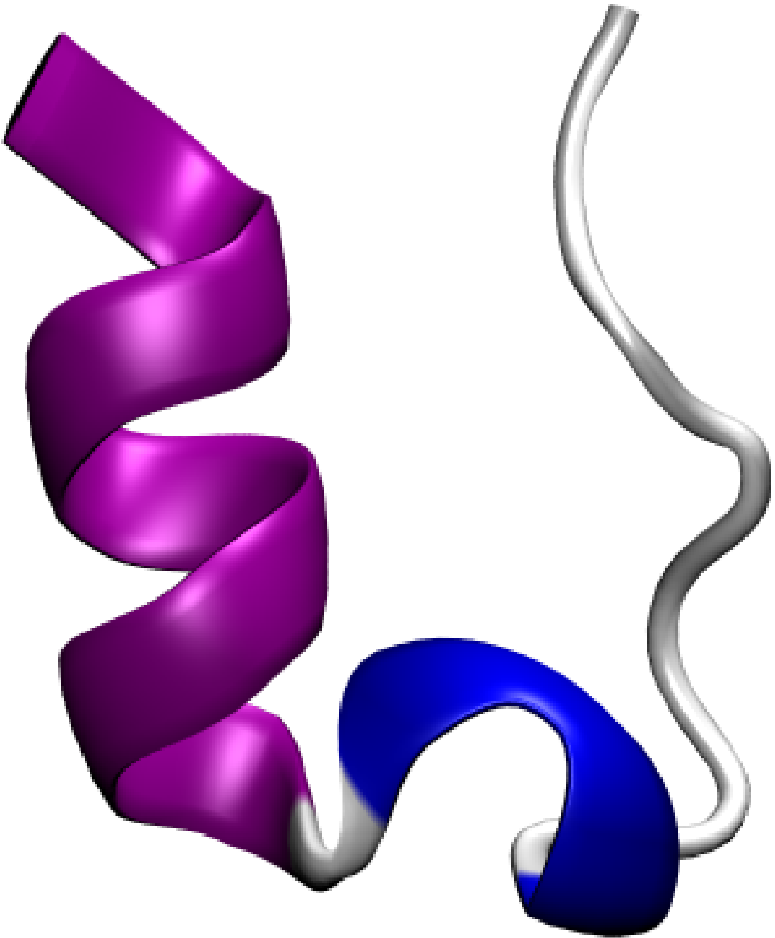}} & \thead{Trp-cage Anton (TC10b) \\ \includegraphics[scale=0.2]{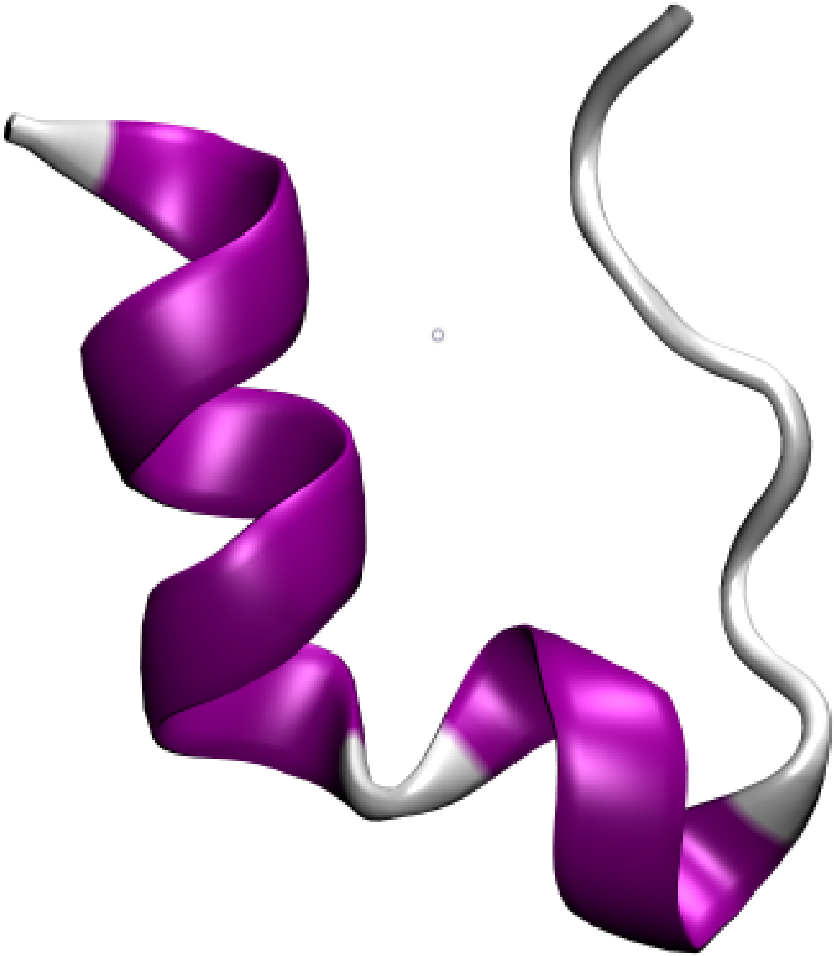}} & \thead{NTL9 \\ \includegraphics[scale=0.2]{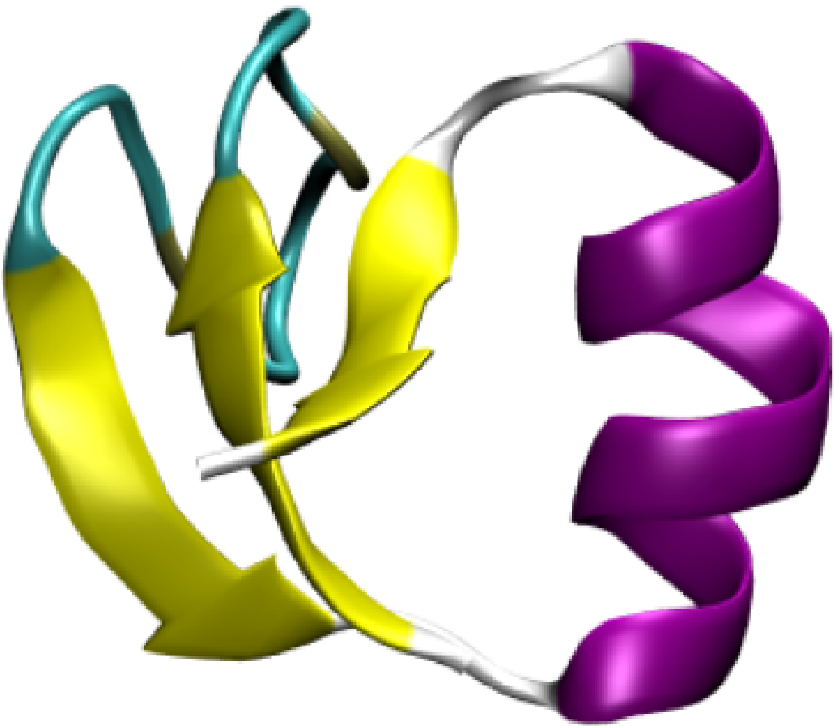}} & \thead{Protein B \\ \includegraphics[scale=0.25]{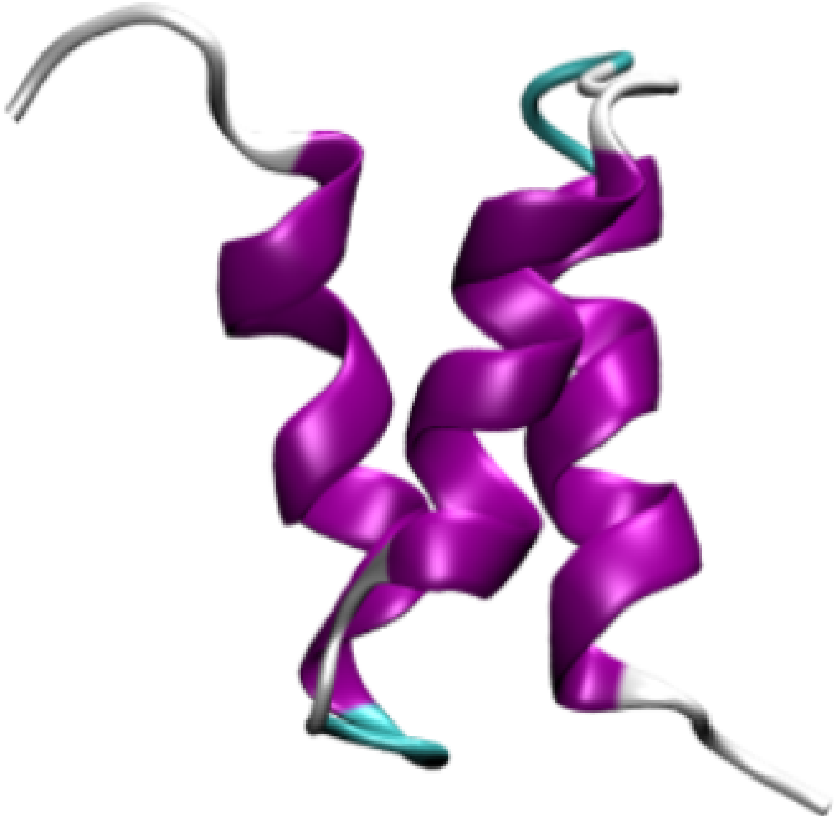}} \tabularnewline
\hline 
\hline 
\makecell{Trajectory length in $\mu$s} & 3.2 & 208 & 1877 & 104 \tabularnewline
\hline 
\makecell{Number of frames} & 1,577,520 & 1,044,000 & 9,389,654 & 520,250\tabularnewline
\hline 
\makecell{Input CVs dimensionality} & 190 & 190 & 703 & 1081\tabularnewline
\hline 
\makecell{Number of cc\_analysis dimensions} & 20 & 20 & 20 & 30\tabularnewline
\hline 
\makecell{Average iteration time \\on our local workstation \\(see SI, Sec.~S-V) {[}min{]}} & 15 & 18 & 55 & 12\tabularnewline
\hline 
\makecell{Average iteration time \\ over all used \\ CPU threads {[}min{]}} & \makecell{24 x 15 \\ = 360} & \makecell{24 x 18 \\ = 432} & \makecell{24 x 55 \\ = 1320} & \makecell{24 x 12 \\ = 288}\tabularnewline
\hline 
\makecell{Frames assigned to clusters \\ after 10 iterations} & 60\% & 33.1\% & 80.9\% & 20\%\tabularnewline
\hline
\makecell{Total CPU time \\ over all iterations {[}min{]}} & 3600 & 4320 & 13200 & 2880\tabularnewline
\hline 
\end{tabular}

\caption{\label{tab:Performance-Table} Proteins analysed in this study and performance overview of the clustering scheme.}

\end{table*}

\section{Results and discussion}

\subsection{Description of the proteins' trajectories used for the analysis}

In order to illustrate the capability and performance of the proposed scheme, we chose four test systems: 40 temperature replica exchange (RE) trajectories of the Trp-cage protein (TC5b) analysed in the original encodermap paper~\citep{encodermap}; the other three systems are long trajectories of Trp-cage (TC10b), NTL9 and Protein B simulated by the Shaw group on the Anton supercomputer~\citep{shaw-trajectories} and generously provided by them. The four systems are listed in \Cref{tab:Performance-Table}. For all the systems we chose distances between C$_\alpha$ atoms as the input collective variables. 
 
The first protein we analyse in this work is the Trp-cage system (TC5b) (Trp-cage RE). It is a comparatively small protein (20 residues) which has a very stable native state when simulated at room temperature. The combination of 40 temperature replica exchange trajectories (temperature range from 300 to 570 K, 3.2~\textmu s  of simulation time, 1,577,520 frames) give a very diverse mixture of structures including trajectories where the system is very stable and barely moves away from the native state, as well as highly disordered trajectories where high-energy conformations are visited. This combination of conformations makes the data set extremely diverse and complicated for the analysis due to the high number of expected clusters with extremely varying size and density. 

\begin{figure*}[!htb]
\begin{centering}
\includegraphics[width=1.0\linewidth]{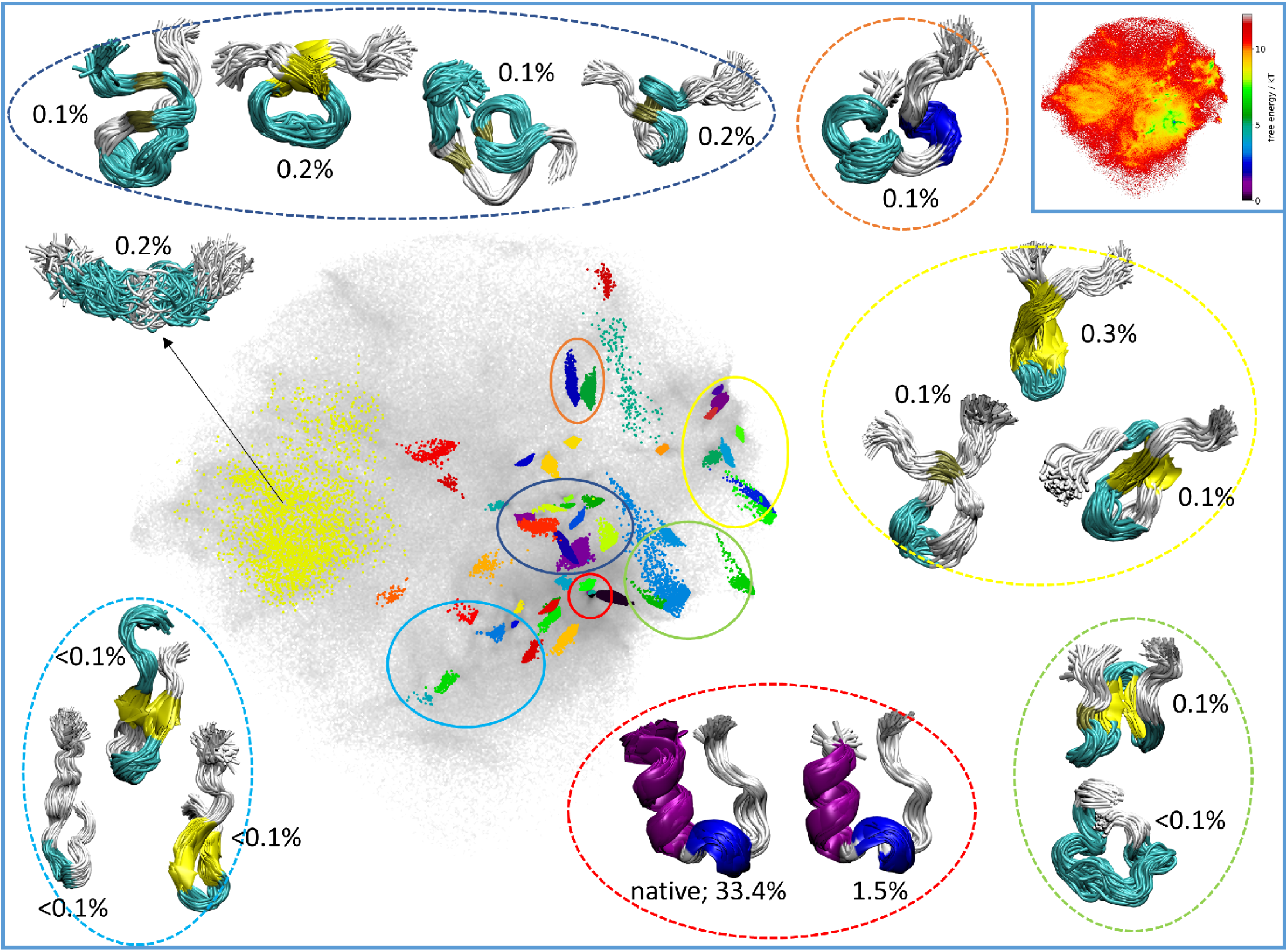}
\caption{\label{fig:trp-cage-results}
Trp-Cage TC5b (40 temperature RE trajectories): 
Exemplary conformations of the most populated clusters
found in each of the areas indicated by coloured circles and their populations in percentages. The cluster representatives show the average secondary structure over the entire cluster. The clusters are coloured randomly, the colours repeat. Therefore clusters that have the same colour but are separated in the 2D space contain different conformations. The depicted clusters hold 36.5\% of all conformations. Most of the remaining 24\% of conformations that have been assigned to clusters are slight variations of the native structure and are not shown here due to visibility reasons. The cluster that is referred to by an arrow is one of the fuzzy clusters that were generated by increasing the RMSD cutoff.
Top right: a histogram of the 2D encodermap space.}
\end{centering}
\end{figure*}

\begin{figure*}[!htb]
\begin{centering}
\includegraphics[width=1.0\linewidth]{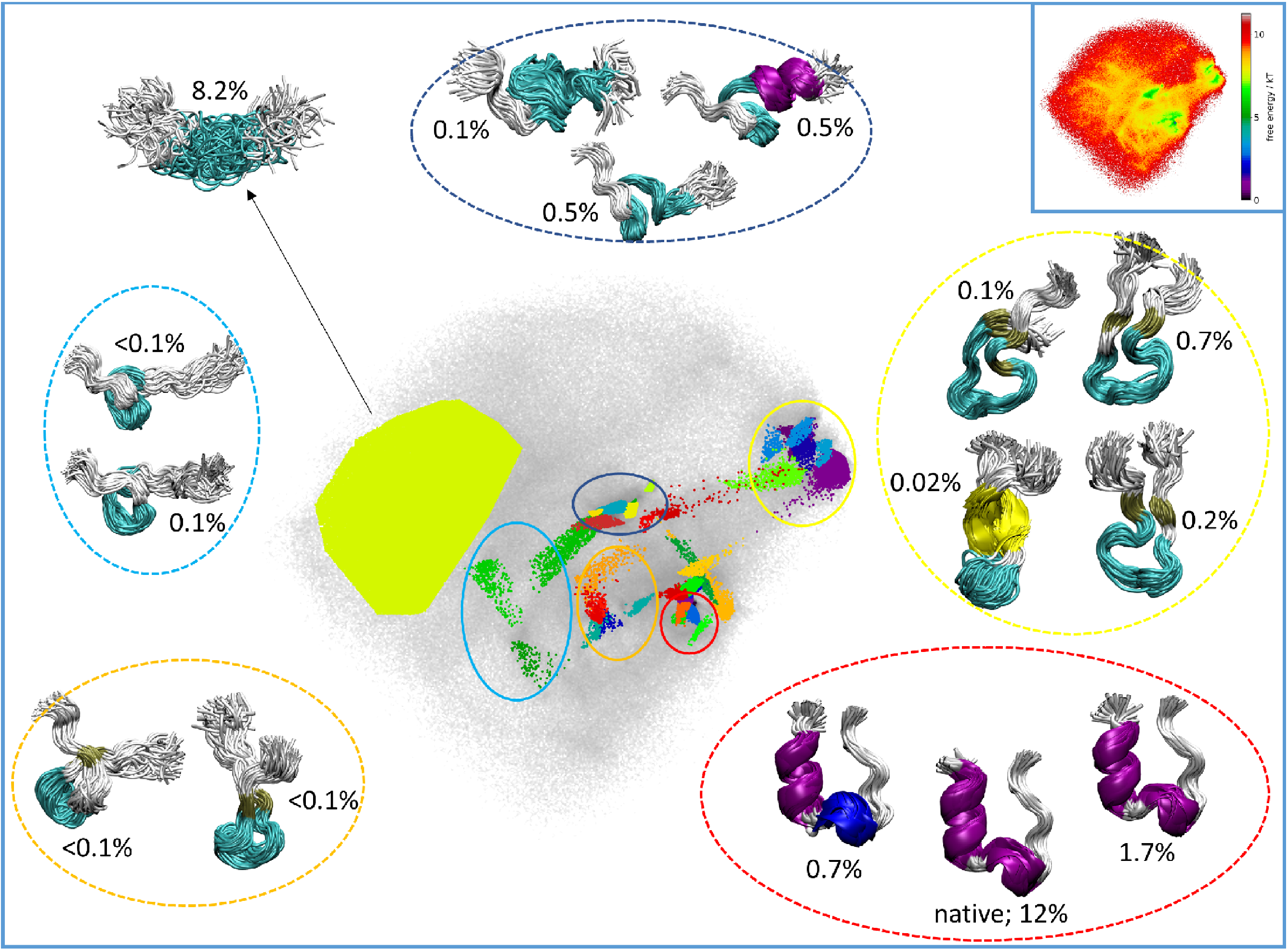}
\caption{\label{fig:trp-cage-results_anton}
The most populated clusters and respective conformations of Trp-Cage TC10b~\cite{shaw-trajectories} projected to the same 2D encodermap space as TC5b (\Cref{fig:trp-cage-results}).bTop right: a histogram of the 2D projection.}
\end{centering}
\end{figure*}

Secondly we consider the K8A mutant of the thermostable Trp-cage variant TC10b (Trp-cage Anton) simulated by \citet{shaw-trajectories} (208 \textmu s; 1,044,000 frames). This simulation was run at 290 K and produced a much more disordered trajectory compared to the low temperature replica simulations of the TC5b system. Despite the fact that TC5b and the K8A mutant of TC10b have slightly different amino acid sequences, we use the same trained encodermap to project both systems in the same 2D map (see \Cref{fig:trp-cage-results} and \Cref{fig:trp-cage-results_anton}), since both systems have the same number of residues and therefore the same dimensionality of CVs. This offers the opportunity to demonstrate that different systems can be compared to each other very nicely when projected to the same 2D space.

Next we probed our clustering scheme with extremely long (1877 \textmu s \footnote{We used the trajectories 0, 2 and 3 according to the nomenclature of Ref.~\citenum{shaw-trajectories}. We have not used trajectory 1 because the topology file for this specific trajectory
differs slightly form the other three in terms of the order and the
numbering of the atoms. This issue has also been reported by other
researchers~\citep{NTL9_issues}.}; 9,389,654 frames) simulations~\citep{shaw-trajectories} of the larger (39 amino acids) N-terminal fragment of ribosomal protein L9 (NTL9) which has an incredibly stable native state. Besides the possibility to show how the algorithm deals with this extremely large data set, the system has also been studied by several other researchers~\citep{pande, VAMPnets}. This allows us to compare our results to their findings.
\citet{pande} reported on very low populated states which involve register-shifts between the residues that are involved in the formation of the beta sheet structures of NTL9. This opens the opportunity to show whether our clustering workflow is able to identify both very large states, as well as extremely lowly populated states in the same data set. 

Lastly we chose to analyse the protein B simulations (104 \textmu s;
520,250 frames) \citep{shaw-trajectories}. Compared to the aforementioned proteins protein B does not have a single very stable state, instead three helices can move quite easily against each other. This leads to a broad conformational space, where the energy barriers between the individual states are very small. Therefore the individual conformational states are not as easily separable and rather fade/transition into each other. Taking into account the long simulation time this system is very hard to cluster conformationally.

To demonstrate how our clustering scheme works we chose to apply it to these four systems that pose very diverse challenges (e.g. an extremely large data set, both highly and very lowly populated states in the same data, differences in the amount of folded/unfolded conformations along the trajectories). For each of the systems we initially conducted the same amount of clustering iterations (10) and then evaluated the resulting clustering and decided whether for a given system additional iterations were needed.

\subsection{Trp-cage}
\label{subsec:trp-cage}

\paragraph{TC5b.} For the RE simulations of the Trp-cage the clustering scheme was run over 10 iterations and assigned 60.5\% of all conformations to clusters. \Cref{fig:trp-cage-results} shows an encodermap projection of all 40 replicas with some of the most populated clusters found after 10 iterations and representative conformations of these clusters. Similar conformations are grouped together and rare structures are spread out across the map. For example, the native conformation
of Trp-cage RE (33.4\% of all conformations) is shown in the bottom right of the 2D map in \Cref{fig:trp-cage-results}. On the bottom left conformations with one turn near the middle of the backbone are located. The two parts of the backbone chain of these conformations lie right next to each other and partially form beta-sheet structures. 

Using a larger cutoff distance in the RMSD-based assignment of structures to the clusters (the other clusters were generated by applying a 1.8~\AA\ RMSD cutoff to the central conformation) we obtained larger and quite diffuse clusters of extended conformations (\changes{one of these clusters is shown in the left part of the projection in \Cref{fig:trp-cage-results} where it is referred to by an arrow). An appropriate size of this RMSD cutoff was defined for each fuzzy cluster individually by computing the mean value of the largest 20\% of the RMSD values between the centroid and cluster members of the cluster identified in the current iteration (it is equal to 5.5~\AA\ for the cluster shown here).
Before we identify fuzzy clusters, we first continuously assign structures based on a fixed RMSD cutoff (1.8~\AA\ for TC5b) until one of the stopping points defined in \Cref{subsec:our_method} is reached (average cluster size for TC5b). Once this stopping point is reached, the RMSD cutoff is adjusted in the way explained above and fuzzy clusters are obtained. Thereby one ensures that all conformations that can be assigned to well-defined clusters are removed from consideration before identifying fuzzy clusters.}
The usage of such a varying cutoff can be very helpful in order to identify diffuse clusters, where the members share a certain structural motif but do not converge to a very defined conformation, just like the cluster shown here.

From the clustering results shown in \Cref{fig:trp-cage-results} one can see that the proposed clustering workflow manages to efficiently identify structurally very well defined clusters for the TC5b system. Over 10 clustering iterations it assigned 60.5\% of all conformations to 260 clusters. Besides the highly populated native state (33.4\%), the algorithm also finds very "rare" states, which contain only a very small amount of conformations ($\leq$0.1\%) but show nevertheless a very defined structural identity.

\paragraph{TC10b.} \Cref{fig:trp-cage-results_anton} shows the same analysis applied to the trajectory of the K8A mutant of TC10b Trp-cage. 
We used the encodermap which we trained on TC5b to project the trajectories to the same 2D space. The identification of clusters however is of course entirely independent and unique for both cases, since the clustering is done in the higher dimensional cc\_analysis space.

Notably, the backbone conformation of the native state of this mutant is extremely similar to the one in the TC5b system. However this biggest cluster only contains 12\% of all conformations along the trajectory compared to the 33.4\% in the case of the TC5b system. If all clusters whose central conformation are within a 2~\AA\ RMSD to the native conformation are combined, we get native conformation percentage of 16.9\%. This is in excellent agreement with the native cluster sizes reported by \citet{trp-cage_levy, trp-cage_ghorbani} who analysed the same Trp-cage trajectories provided by \citet{shaw-trajectories}. Furthermore our 33.4\% of assigned conformations coincide very well with the reporting of \citet{trp-cage_ferguson}. They found a total of 31\% of conformations distributed over eight metastable macrostates and the remaining 69\% as one big "molten globule" state. 

The TC10b trajectory is more disordered, this can be seen by the more homogeneous projection in 2D space (upper right plot in \Cref{fig:trp-cage-results_anton}) and the RMSD values to the native conformation in SI, Sec.~S-III, Fig.~S3. This is also the reason why the clustering scheme assigned only 33.4\% of all conformations to clusters after 10 iterations. If more frames should be assigned to clusters, more clustering iterations can be performed, the RMSD cutoff can be increased or both can be done simultaneously (for the Protein B system we show the results of this approach later in the article). 

However the clusters in the very center of the map (dark blue circle) are much more compact and collapsed compared to the clusters that were found in the similar area of Trp-cage RE's 2D projection. Also some of the clusters that were found in the very bottom of the left hand side of the map in case of the replica trajectories (light blue circle) were not found at all in the TC10b trajectory. The very large and diffuse cluster on the left side of the map is present in both systems as well.

\begin{figure}
\begin{centering}
\includegraphics[width=1.0\linewidth]{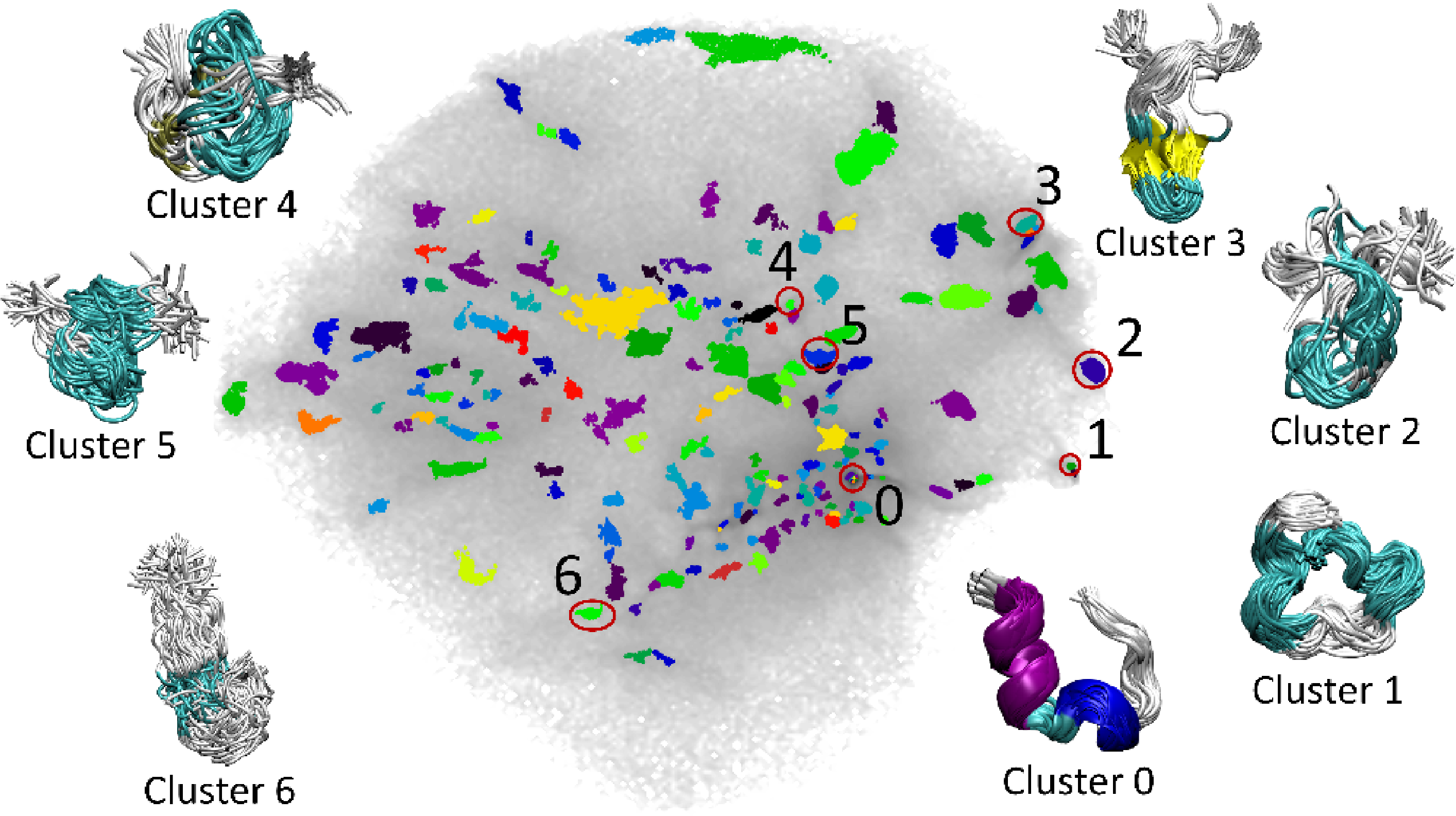}
\caption{\label{fig:cluster_comparison_encodermap+hdbscan}
2D encodermap space of TC5b clustered with HDBSCAN. Representations of chosen clusters that have the same location in the 2D map as clusters found with the clustering scheme in \Cref{fig:trp-cage-results} are shown.}
\end{centering}
\end{figure}

\paragraph{Clustering directly in 2D space of TC5b.} The clustering discussed above was done in a 20 dimensional space after applying the cc\_analysis algorithm and only displayed at a 2D projection done with encodermap. In order to demonstrate the advantages of our approach we also directly clustered the 2D encodermap space using the HDBSCAN. The encodermap space that we used for this clustering is the same space that we used to visualize the cc\_analysis clustering in \Cref{fig:trp-cage-results} and \Cref{fig:trp-cage-results_anton}. The results of this clustering and a few chosen clusters can be seen in \Cref{fig:cluster_comparison_encodermap+hdbscan}. In total this clustering assigned 13.5\% of all conformations to 362 clusters. The biggest cluster that was found is the native cluster, however it only contains 0.8\% of all conformations compared to the 33.4\% that were found by clustering the cc\_analysis space.
The clustering in the 2D space identifies some structurally very well defined clusters, such as the clusters 0, 1 and 3, but also a lot of very diffuse and inhomogeneous clusters. To quantify this inhomogeneity we computed the average of the internal cluster RMSDs. For the TC5b system our clustering workflow resulted in an average cluster RMSD of 1.34~\AA\ and a weighted average RMSD of 1.03~\AA, where weights are defined as the fraction of each cluster to all clustered data. The average RMSD for the direct clustering in the 2D space is 2.25~\AA\ and the weighted average RMSD is 2.73~\AA. This clearly shows that the internal cluster RMSD variance is on average much larger when clustering directly in the 2D space. Furthermore the clustering in the 2D space itself naturally highly depends on the quality of the 2D map.

Other than the much clearer conformational identity of the individual clusters (shown via internal cluster RMSDs), our clustering scheme also manages to assign 60.5\% of all conformations to different clusters. Compared to that the clustering in the 2D projection only assigned 9-14\% of all conformations depending on the choice of clustering parameters.

\paragraph{Comparison to other clustering approaches.} For a further assessment of our clustering scheme we have also applied a frequently used clustering routine to the TC5b data. In Si, Sec.~S-IV and Figs.~S4 and S5 the results of applying the k-means algorithm to an 11 dimensional PCA projection of the same CVs (pairwise C$_\alpha$ distances of TC5b) are shown.

In summary, the scheme identified both structurally very defined as well as quite diffuse clusters in considered systems. Even though the combination of the 40 RE trajectories produces a very diverse data set, the clustering scheme manages to assign a large amount of the conformations to clusters (60\%).
Our clustering results for the TC10b are in a very good agreement with the findings of other researchers~\citep{trp-cage_levy, trp-cage_ghorbani, trp-cage_ferguson}.
Furthermore the comparison to a clustering in the 2D space clearly shows the superiority of using more dimensions obtained with the cc\_analysis algorithm in HDBSCAN over just relying on a low-dimensional representation alone.
\begin{figure*}[ht]
\begin{centering}
\includegraphics[width=1.0\linewidth]{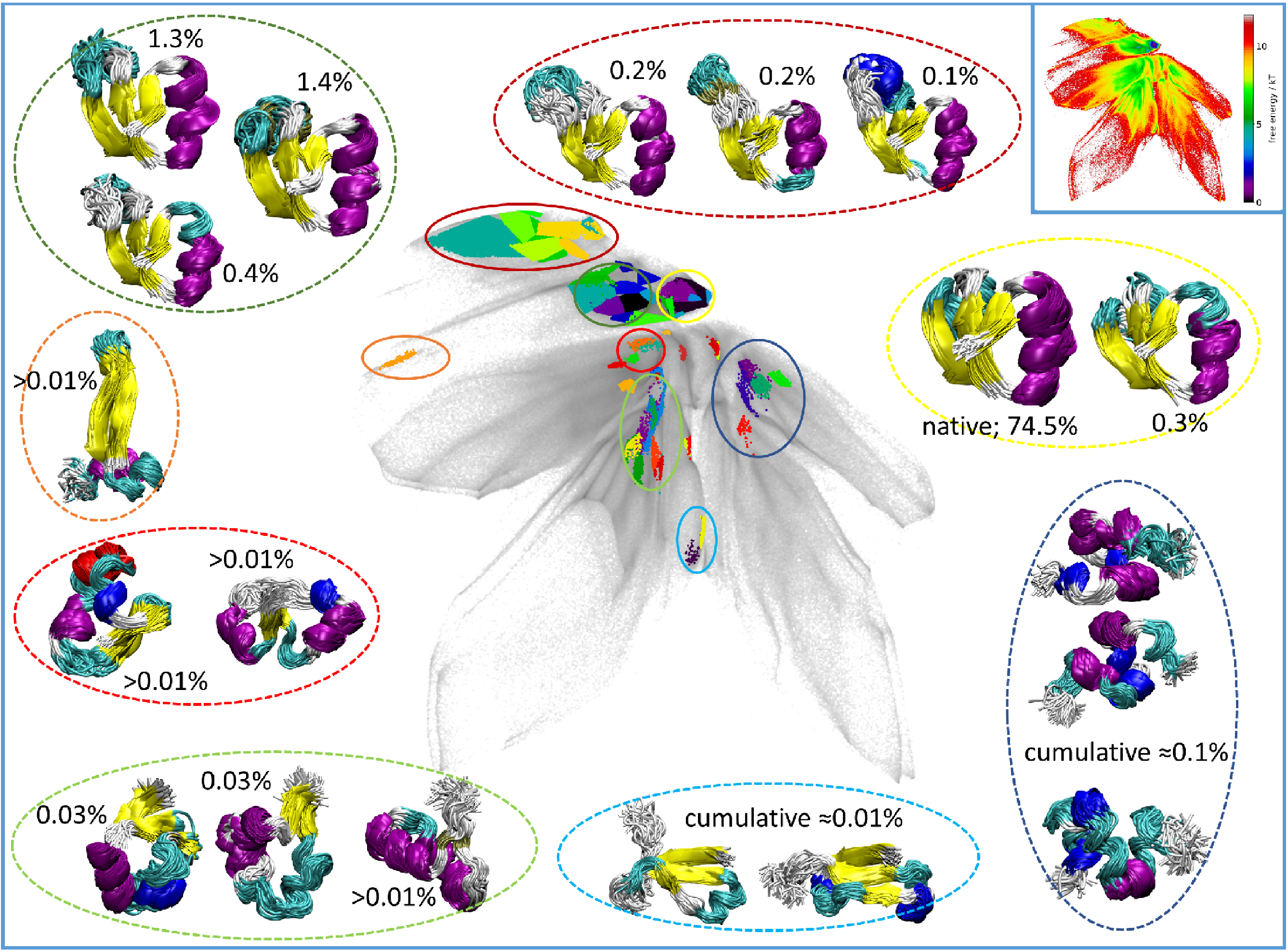}
\caption{\label{fig:NTL9 results}
The 2D encodermap projection of NTL9. The projection can be approximately divided into three parts: the upper part with the most dense areas (where the native-like states are located); the lower left and right planes divided by an unpopulated vertical gap. The left side includes various conformations with a singular beta sheet formed mostly between the beginning and the end of the protein. In contrast on the right side lie mostly extended conformations with multiple helices along the backbone.
Exemplary conformations of some of the most populated clusters found in each of the marked areas in the map and their populations are shown. All clusters in the yellow circle are extremely similar to the native cluster and can be summed up to a total of ~76\% of all conformations. The structures that are shown here make up ~78.4\% of all conformations.  Top right: Histogram of the 2D encodermap space.}
\end{centering}
\end{figure*}
\subsection{NTL9}

Next we examined very long (1877 \textmu s) simulations of NTL9~\cite{shaw-trajectories}. With 9.38 million frames to cluster, this system is an ideal candidate to demonstrate how
the proposed algorithm copes with large amounts of data. After 10 iterations 81\% of all conformations were assigned to clusters.
\Cref{fig:NTL9 results} shows a 2D projection made with encodermap, where points are colored according to the clusters found after ten iterations of the scheme and a histogram of the 2D space in the upper right corner.  In total we found 157 clusters and assigned them 81\% of all conformations over 10 clustering iterations.

A comparison of the timeseries of the RMSD values to the folded state to the respective data of the Trp-cage Anton simulations (SI, Sec.~S-III, Fig.~S3) reveals that the two systems exhibit very different dynamics. While in the Trp-cage case the RMSDs show the disordered nature of the system, in
the case of the NTL9 trajectories the RMSDs are predominantly quite
low and only spike up to larger values for rather short time periods.
This suggests that the NTL9 system resides in a native-like state
for the majority of the simulated time. This is confirmed during the very first iteration of the clustering scheme. There we found two clusters which make up for 75.8\% of all conformations.

This example also nicely illustrates how the iterative clustering
approach can be efficient in identifying clusters of very different size and density (highly populated native states and low populated clusters).
After finding and removing the first two clusters (75.8\% of the data) the clustering algorithm becomes much more sensitive towards the less
dense areas in the CV-space in the following clustering iterations.

\begin{figure}
\begin{centering}
\includegraphics[width=1.0\linewidth]{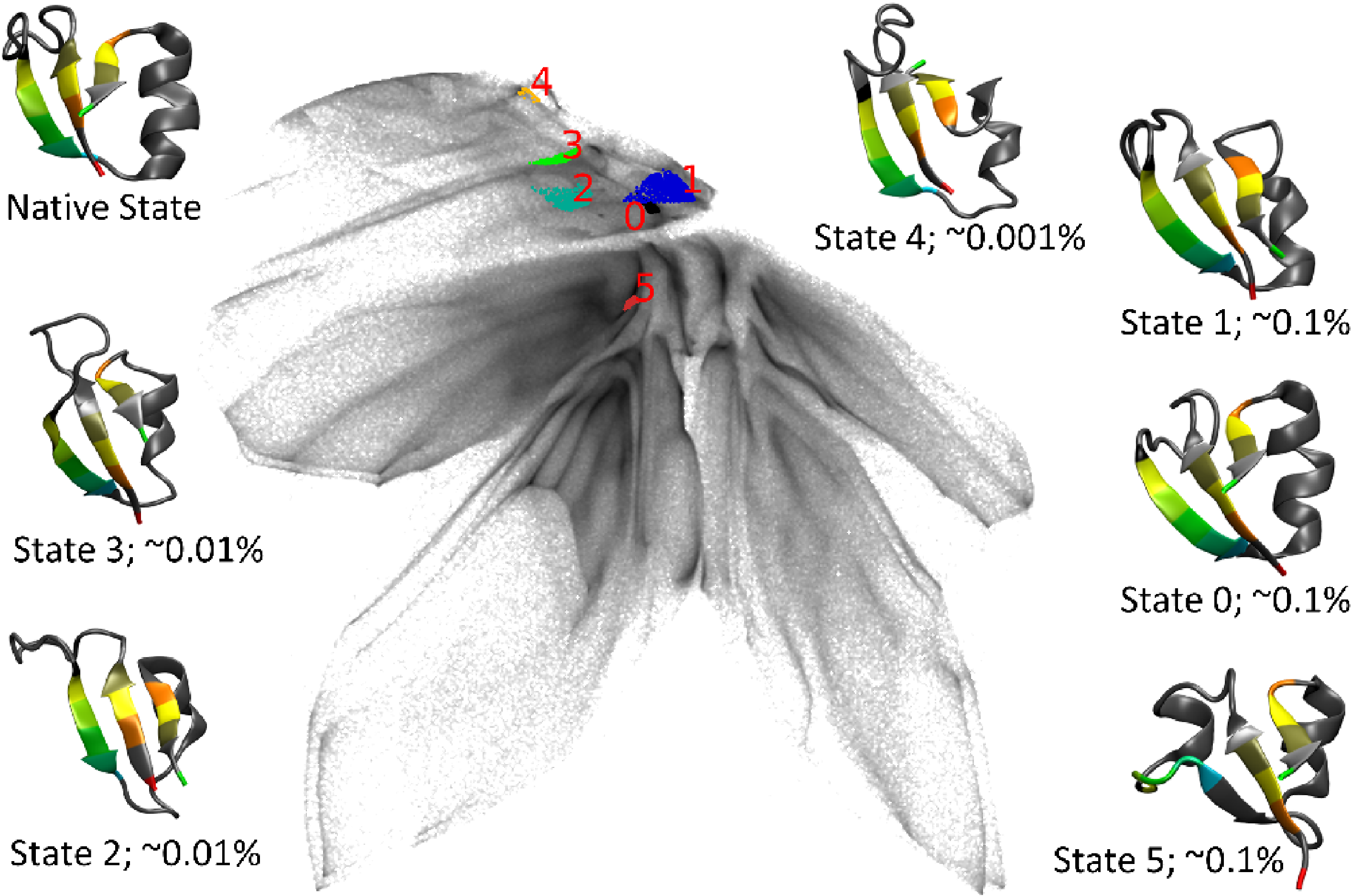}
\caption{\label{fig:Register-shifted-states}
Register-shifted states found in the NTL9 trajectories 0, 2 and 3. The residues which form the beta
sheets in the native state are colored based on their residue ID.
}
\end{centering}
\end{figure}

We compared our clustering results with other publications analyzing the NTL9 trajectories from Ref.~\cite{shaw-trajectories}. \citet{VAMPnets} applied the VAMPnets to trajectory~0 and found in total 89.1\% of folded, native like conformations. If we take the clusters we found by analysing the trajectories 0, 2 and 3 and evaluate the conformations stemming from trajectory 0 (trajectory 0 resides in the native-like state for a larger fraction of the simulated time; see RMSD plots in SI, Sec.~S-III, Fig.~S3, the amount of folded, native-like conformations we find is in very good agreement with \cite{VAMPnets}. 
Furthermore \citet{pande} reported the finding of three ``register-shifted'' states, which are very low populated and therefore very hard to find. ``Register-shifted'' refers to the identity of the specific residues involved in forming the beta sheet structure in the native-like states (residues 1-6, 16-21 and 35-39).  With our method we identified six different register-shifted states in the NTL9 trajectories 0, 2 and 3 (see \Cref{fig:Register-shifted-states}).

The states~0, 1 and 2 are the ones which were also found in \cite{pande}. 
To our knowledge states~3, 4 and 5 have not been reported yet. 
In state~0 the central of the three beta-sheet strands is shifted downwards, whereas in state~2 the rightmost strand is shifted downwards. In state~1 both the middle and the rightmost strands are dislocated compared to the native state.
State~3 is similar to state~1 in the fact that both the middle and the rightmost strands are shifted, however in state~3 the rightmost strand is shifted upwards and not downwards like in state~1. 
Among these six states state~4 is unique since there the rightmost strand is turned by 180 degrees. 
Finally state~5 differ from other states in having an extra helix along the chain between the leftmost and the middle strand. Because of this additional helix the leftmost strand is extremely shifted compared to the native state. 

The identification of these register-shifted states highlights one asset of the proposed workflow. It is able to find both very large states (native, 74.5\%) as well as very low populated clusters (\textless0.001\%) in the same data set.

\subsection{Protein-B}

The last system we analysed is Protein B. 
This system does not have a very stable native state, instead the three helices can move against each other relatively freely. This can be seen in the timeseries of the 
RMSD to the closest experimental homologue (1PRB) shown in SI, Sec.~S-III, Fig.~S3. There are no extended periods where the values are stable over some time, meaning there are no large free-energy barriers separating the various accessible conformations and thus the system constantly transitions into different conformations. 
This has also been found in \cite{shaw-trajectories}, where authors stated that they were unable to identify a free-energy barrier between folded and unfolded states for Protein B (tested over many different reaction coordinates). 

\begin{figure*}[!htbp]
\begin{centering}
\includegraphics[width=1.0\linewidth]{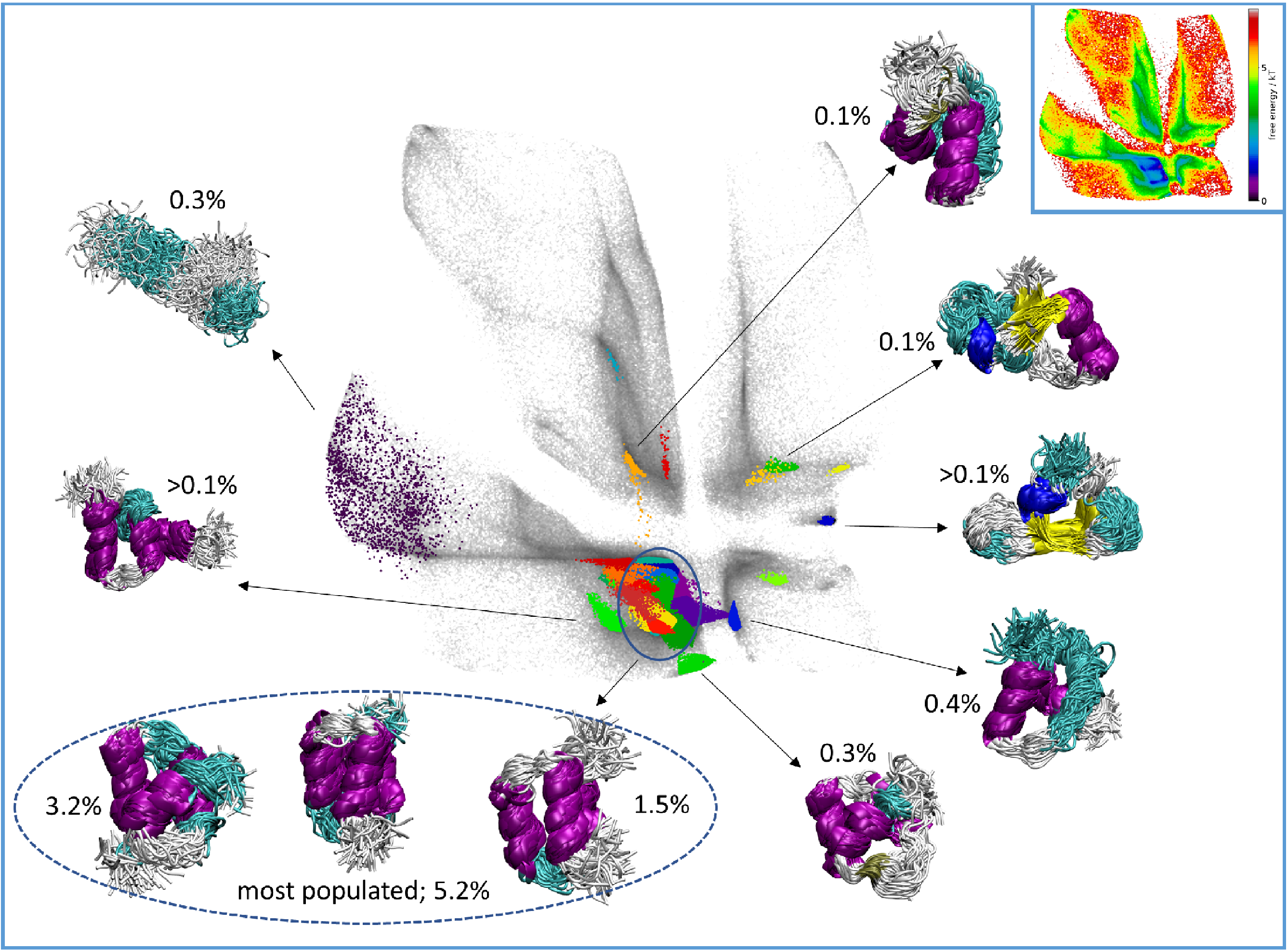}
\caption{\label{fig:PRB-results}
Protein B: 
Exemplary conformations of some of the most populated clusters
found for the Protein B system after 10 clustering iterations and their populations;
Top right: Histogram of the 2D encodermap space.}
\end{centering}
\end{figure*}

Such a highly dynamic system is very challenging for a conformational clustering. Here we want to show where our algorithm has its limitations and what can be done to get a satisfactory clustering result. 
\Cref{fig:PRB-results} gives an overview of some of the clusters found after ten iterations of the scheme. 
These clusters include only 20\% of the Protein B trajectory and thus 80\% of all conformations are still unclustered. 

In order to have more data assigned to clusters two parameters can be adjusted. 
First, the RMSD cutoff value can be increased and thereby more conformations can be assigned to the found clusters. 
In this specific case this adjustment is justified, since due to the low free-energy barriers between different states, the individual clusters are not as sharply defined in terms of their conformations. 
In the 10 clustering iterations which are shown in \Cref{fig:PRB-results} we used a RMSD cutoff of 3.0~\AA. In a second run we increased it to 3.5~\AA. This resulted in an assignment of 31\% of all conformations to generally more loosely defined clusters.

A second approach is to increase the amount of clustering iterations. 
For the first ten clustering iterations of previously analysed systems, we tuned the clustering parameters manually. This includes the choice of the number of cc\_analysis dimensions, as well as the min\_ samples and min\_cluster\_size parameters of HDBSCAN. 
However such a manual adjustment of the parameters is of course not feasible for automating the script in order to perform many more clustering iterations. Since the amount of cc\_analysis dimensions needs to be very rarely changed once a suitable amount has been identified in the first clustering iteration, the automation of the script only relies on the choice of the HDBSCAN parameters.
Once the amount of clusters found in a single iteration falls below a certain threshold (10 clusters in this case), the numerical values of the min\_samples and min\_cluster\_size parameters of HDBSCAN are slightly decreased. This leads to the detection of smaller clusters that have not been identified before. By applying this automation approach after the first 10 iterations to Protein~B and using a RMSD cutoff of 3.5~\AA,
we could assign 44.3\% of all conformations to clusters over 100 iterations, which took roughly 15 hours on our workstation. 

\section{Discussion}

The Trp-cage system (TC5b) is a relatively small protein which has a quite
stable native conformation. The combination of 40 temperature RE trajectories however gives a very diverse data set including (under standard conditions) very improbable high-energy conformations.
Over ten iterations the algorithm managed to assign 60.5\% of all
conformations to clusters, which took on average 360~min per
iteration over all CPU threads (15~min per iteration on a standard
office machine with 24~CPU threads). \Cref{tab:Performance-Table} shows the clustering performance for the four systems discussed here.
By switching the generally static
RMSD cutoff to a varying cutoff we could show that the algorithm can
both generate conformationally very defined clusters as well as quite
diffuse. The conformations assigned to such loose clusters
share a general structural motif. 
The ability to identify both of these cluster types is one of the advantages of the proposed algorithm. 
Furthermore we demonstrate that the clustering workflow is able to directly compare different systems (even if they slightly differ structurally), by projecting them to the same 2D map using the encodermap algorithm. This enables a direct and visual comparison of the sampled phase-spaces of different trajectories and their respective identified states. 
By comparing the clustering result where the clustering is done in a 20-dimensional cc\_analysis space and then projected to a two-dimensional space to a clustering where the clusters are purely found in a 2D encodermap space, we prove an advantage using more dimensions and combine cc\_analysis with encodermap. 
The scheme created clusters with a much clearer structural identity (lower RMSD variance), while being much less dependent on the quality of the 2D map.

We analysed long (9.38~million frames) trajectories of NTL9 to show how the proposed scheme copes with very large amounts of data. On average the algorithm needed 1320~min of computation time
over all CPU threads per iteration (55~min per iteration on our office
machine). Since this system also has one hugely populated native-state, it is also a nice example to demonstrate an advantage of the iterative
clustering. After the clusters with the native states are removed from consideration, the algorithm becomes much more sensitive towards less populated areas
in the following iterations. Applying this approach we could identify three very low populated register-shifted states, which have been reported before~\cite{pande}, and three not yet seen register-shifted states. 

Lastly we looked at is Protein B, which is a highly dynamic
system. To analyse this 1.04~million frames trajectory it took on average 288~min of computation time per iteration (12~min per iteration on our office machine). 
This system has no large free-energy barriers separating the various conformations, which makes it very difficult to cluster.
This was confirmed by the fact that after ten clustering iterations
only 20\% of all conformations could be assigned to clusters. However
by increasing the RMSD cutoff from 3.0~\AA\ to 3.5~\AA\ we could already
increase the amount of assigned conformations to 31\%, which of course
resulted in slightly less structurally defined clusters. It is also possible to automate the clustering and run until a certain amount of conformations are assigned
to clusters or until a given number of iterations is reached. In this
specific case we ran the scheme for 100 automated iterations ($\approx$15 hours), during which 44.3\%
of the conformations were assigned to clusters. 

For all considered systems the proposed workflow was able to identify defined clusters at the cost of leaving some amount of the trajectories unassigned. As we have shown here, the rest of the structures does not belong to any specific clusters and can be considered as unfolded or transition states. We intentionally do not propose any additional steps to assign or classify those conformations as it is highly dependant on the intended application of the data. For example in case the data is used to build subsequent kinetic models the rest of the points can be assigned to the nearest (e.g. in simulation time) cluster using methods such as PCCA+ analysis~\cite{Deuflhard2005}, or defined as a metastable transition state as in Ref.~\citenum{trp-cage_ferguson}. It can also be defined as noise and used as discussed in Ref.~\citenum{Keller2016}.

All performance data is shown in \Cref{tab:Performance-Table} and was obtained
by running the clustering scheme script on the office workstation described
in SI, Sec.~S-V. The proposed workflow is, however, highly parallelizable, since the computationally most expensive step is the assignment of additional data points to the initially identified clusters in the small subset based on the convex hull and the RMSD criterion. If a large amount of CPU cores are available, the 2D encodermap projection array can be split by the amount of cores and the assignment can thereby be run in parallel which leads to a significant speed up. 

The convex hull around the clusters identified in the small subset is used to reduce the amount of RMSD computations that have to be performed when assigning additional conformations in each clustering iteration. This however might in principle lead to the exclusion of data points that might otherwise have been assigned to some of the clusters. In order to get an idea of the magnitude of this ``loss'' of potential cluster members, we computed the RMSD of all data which was labeled as noise (623,000 conformations; 39.5\%) to each of the cluster centers of TC5b (260 clusters). This computationally very expensive task took an additional 5~hours on our working machine. We found that 42,000 conformations (2.7\%) were not assigned to the identified clusters due to the convex hull criterion. When keeping in mind that the entire 10 iteration clustering process took 2.5~hours, the "loss" of 2.7\% of unclustered data can be considered a worthy trade-off.

Another point to consider is that due to the convex hull criterion clusters can be split. If data points that would be assigned to a certain cluster by reason of the RMSD criterion lie outside of the convex hull, they could be identified as another cluster in one of the following clustering iterations. In such cases it can make sense to merge these clusters in hindsight, due to their very similar structural identity. In order to showcase such a merge, we again analysed TC5b. We computed the RMSDs between all of the 260 central cluster conformations and merged all clusters that had a RMSD of $\leq$ 1~\AA. This resulted in a reduction to 201 clusters with only very marginal influence on the average internal cluster RMSDs.

The code for the encodermap algorithm is available on the following github page  \url{ https://github.com/AG-Peter/encodermap}.
The cc\_analysis code can be found under \url{https://strucbio.biologie.uni-konstanz.de/xdswiki/index.php/Cc_analysis}.

\section{Conclusion}

We developed a clustering scheme which combines two different dimensionality reduction algorithms (cc\_analysis and encodermap) and the HDBSCAN in an iterative approach to perform fast and accurate clustering of molecular dynamics simulations' trajectories. The cc\_analysis dimensionality reduction method was first applied to protein simulation data. The method projects collective variables to a usually relatively high-dimensional ($\sim$10-40 dim) unit sphere, separating noise and fluctuations from important structural information. Then the data can be efficiently clustered by density based clustering methods, such as HDBSCAN. The iterative application of HDBSCAN allows to account for the inhomogeneity in population and density of the projected points, which is very typical for protein simulation data. As cc\_analysis relies on the calculations of correlation matrices between each frame, this drastically limits the amount of data one can project simultaneously. To allow processing of long simulation trajectories we included encodermap to the scheme. In addition to the obvious advantage of the two-dimensional visualisation it is used -- in combination with a RMSD-based acceptance criterion -- for a fast structure-based assignment of additional points to the clusters initially identified in the higher dimensional projection done with cc\_analysis. To demonstrate the accuracy and performance of the proposed scheme we applied the clustering scheme to four test systems: replica exchange simulations of Trp-cage and three long trajectories of a Trp-cage mutant, NTL9 and Protein B generated on the Anton supercomputer. By applying the scheme to these four test systems we could show that: the algorithm can efficiently handle very large amounts of data, that it can be used to compare the clusters of structurally different systems in one 2D map, and that it can also be applied to cluster systems which do not have very stable native states and are therefore intrinsically very difficult to cluster conformationally. 
Furthermore the algorithm is able to find clusters independent of their size. By varying a RMSD cutoff both conformationally very well defined clusters, as well as fuzzy clusters, whose members only share an overall structural motive, can be identified. 


\section{Supporting information}
Supporting Information (PDF) includes:

(S-I): Methods to chose parameters for cc\_analysis and HDBSCAN.

(S-II): Stopping criteria for the clustering workflow.

(S-III): RMSD plots of trajectories for Trp-cage, Protein B and NTL9.

(S-IV): Comparison of the proposed clustering workflow to PCA and k-means clustering for Trp-cage (TC5b).

(S-V): Workstation specifications.


\section{Acknowledgements}

This work was supported by the DFG through CRC~969.
We also greatly appreciate the computing time on bwHPC clusters which was used to produce the Trp-cage TC5b trajectories.
Furthermore we would like to thank the D.E. Shaw research group for providing the Trp-cage, NTL9 and Protein B trajectories.


%

\end{document}